\documentclass[prd,aps,float,twocolumn,showpacs,color]{revtex4}
\usepackage{amsmath}
\usepackage{amssymb}
\usepackage{mathrsfs}
\usepackage{natbib}
\usepackage{dcolumn}
\usepackage{float}
\usepackage{graphicx}

\begin{document}
\title{\large \bf Exact general relativistic lensing versus thin lens approximation: the crucial role of the void }
\author{M. Parsi Mood}
\affiliation{Department of Physics, Sharif University of Technology,
Tehran, Iran} \email{parsimood@physics.sharif.edu}

\author{Javad T. Firouzjaee}
\affiliation{School of Astronomy and Physics, Institute for Research in Fundamental Sciences (IPM), Tehran, Iran}
\email{j.taghizadeh.f@ipm.ir}
\author{Reza Mansouri}
\affiliation{Department of Physics, Sharif University of Technology,
Tehran, Iran and \\
  School of Astronomy, Institute for Research in Fundamental Sciences (IPM), Tehran, Iran}
\email{mansouri@ipm.ir}

\date{\today}

\begin{abstract}
We have used an exact general relativistic model structure within a FRW cosmological background based on a LTB metric to
study the gravitational lensing of a cosmological structure. The integration of the geodesic equations turned out to be a
delicate task. We realized that the use of the rank 8(7) and 10(11) Runge-Kutta numerical method leads to a numerical effect
and is therefore unreliable. The so-called semi-implicit Rosenbrock method, however, turned out to be a viable
integration method for our problem. The deviation angle calculated by the integration of the geodesic equations
for different density profiles of the model structure was then compared to those of the corresponding thin
lens approximation. Using the familiar NFW density profile, it is shown that independent of the truncation details
the thin lens approximation differ substantially from the exact relativistic calculation. The difference in
the deflection angle for different impact parameters may be up to about 30 percent. However, using the
modified NFW density profile with a void before going over to the FRW background, as required by an exact
general relativistic model, the thin lens approximation coincides almost exactly with the general relativistic
calculation.

\end{abstract}
\maketitle
\par

 The thin lens (Th-L) approximation in the gravitational lensing is the prevailing method to estimate cosmological
parameters and the mass of large scale structures leading to dark matter and dark energy contents of the universe {\cite{GLenses}, \cite{hoekstra}.
The current view is that this Th-L approximation is accurate enough at the cosmological scales where we are faced with
very weak gravitational fields and potentials. There has already been attempts to compare the Th-L approximation with the
integration of null geodesics in a perturbed cosmological background (\cite{Sas93,Fut95,FriKling11}, see
also \cite{FriKling11} and the references there). However, a full general relativistic calculation based on an exact model is still
missing. There are two sources of misinterpretation of astrophysical phenomenon in a weak gravity environment, depending
on the local or quasi-local phenomena under consideration. In the case of local phenomena the familiar perturbation
theories maybe valid to some extend. There are already detailed studies on this subject (see \cite{Wald1}, \cite{Wald2}, \cite{Wald3}).
However, if quasi-local phenomena or structures come into play we may encounter counter-intuitive effects not detected in
the perturbational approach to the weak field limits. The definition of quasi-local mass in general relativity is one of
these issues which has been extensively studied in general relativity \cite{Szabados}. We have already shown
numerically how different various quasi-local mass definitions of a general relativistic structure may be \cite{taghizadeh}.
Another quasi-local effect relevant to the gravitational lensing is how a spherically symmetric structure is
matched to a FRW background. Such a general relativistic matching is only possible through an underdensity region or a
void \cite{khakshournia}; a fact not realized in the post-Newtonian approaches or cosmological perturbations relevant to
lensing, and missed in all studies comparing the Th-L approaches to a more exact general relativistic lensing calculation.\\
\par
We are interested in the exact general relativistic lensing by an exact solution of Einstein Equations representing
a cosmological structure defined by a spherically symmetric overdensity structure within a FRW universe. There is already
an exact general relativistic model structure within an FRW universe based on a Lema\^{\i}tre, Tolman and Bondi (LTB)
metric \cite{Lem97,Tol34,Bon47} representing an inhomogeneous cosmological model with a structure at its center\cite{taghizadeh}.
Choosing such a model for an extended spherical lens, we study the gravitational lensing in a dynamical cosmological background
in the framework of general relativity by integrating numerically the null geodesic equations to obtain the deflection angle. The
result is then compared with the corresponding Th-L approximation to understand the accuracy of this technology and its possible flaws
in interpreting the structure and the mass of cluster of galaxies. The effect of the cosmological constant in the lensing is
negligible in small scales we are considering \cite{ishak} and only effect the cosmological distances which we will take
into account. That is why we have neglected the cosmological constant in our exact model to avoid unnecessary complexities\\
Take a spherically symmetric cosmological structure in a FRW matter dominated universe with the density $\rho(r,t)$.
This is modeled by a LTB solution of the Einstein equations which is written in the comoving coordinates as ($G=1,c=1$)
\begin{equation}\label{ltbm}
 ds^{2}=-dt^{2}+X^2(r,t)dr^{2}+R^2(t,r)d\Omega^{2}.
\end{equation}
satisfying
\begin{eqnarray}
\rho(r,t)&=&\frac{M'(r)}{4\pi R^{2}R'},\\
X&=&\frac{R'}{\sqrt{1+E(r)}},\\
\dot{R}^{2}&=&E(r)+\frac{2M(r)}{R}.\label{fieldeqn}
\end{eqnarray}
Here $M$ and $E$ are integrating functions, where dot and prime denote partial derivatives with respect to the coordinates $t$
and $r$ respectively. Equation (\ref{fieldeqn}) has three different analytic solution, depending on the value of $E$. The
solution for negative $E$ we are interested in is given by
\begin{eqnarray}\label{ltbc}
R&=&-\frac{M}{E}(1-\cos\eta),\nonumber\\
\eta-\sin\eta&=&\frac{(-E)^{3/2}}{M}(t-t_{b}(r)).
\end{eqnarray}
The solution has three free functions: $t_{b}(r)$, $E(r)$, and $M(r)$. Given that the metric is covariant under the
rescaling $r\rightarrow\tilde{r}(r)$ one of these functions may be fixed.\\
The geodesic equations may be written in the arbitrary plane of $\theta=\frac{\pi}{2}$ due to the spherical symmetry:
\begin{flalign}\label{ltbgeo}
&t:\frac{d^2t}{d\lambda^2} + X\dot{X}\left(\frac{dr}{d\lambda}\right)^2 + R\dot{R}\left(\frac{d\phi}{d\lambda}\right)^2 = 0, \\
&r: \frac{d^2r}{d\lambda^2} + 2\frac{\dot{X}}{X}\frac{dr}{d\lambda}\frac{dt}{d\lambda} + \frac{X'}{X}\left(\frac{dr}{d\lambda}\right)^2-\frac{RR'}{X^2}\left(\frac{d\phi}{d\lambda}\right)^2=0, \\
&\phi: \frac{d^2\phi}{d\lambda^2} + 2\frac{\dot{R}}{R}\frac{dt}{d\lambda}\frac{d\phi}{d\lambda} + 2\frac{R'}{R}\frac{dr}{d\lambda}\frac{d\phi}{d\lambda}=0, \label{geophi}
\end{flalign}
where $\lambda$ is an affine parameter. Equation (\ref{geophi}) expresses the conservation of the angular momentum:
\begin{equation}
L=R^2\frac{d\phi}{d\lambda}=Const.
\end{equation}
We are interested in the light-like geodesics. From the metric we obtain the light-like condition in the form
\begin{equation}\label{null}
\left(\frac{dt}{d\lambda}\right)^2=X^2 \left(\frac{dr}{d\lambda}\right)^2 + R^2 \left(\frac{d\phi}{d\lambda}\right)^2
\end{equation}
 These partial non-linear differential equations can not be solved analytically. To integrate them numerically one has to
specify the three functions $M(r), t_{b}(r)$, and $E(r)$ and all derivatives of the metric functions, using a procedure
proposed in \cite{KH01,BKCH}. We start with a generic density profile and specify it at two different times $t_1, t_2$ as
a function of the coordinate $r$. Now, the numerical procedure is based on the choice of $r$-coordinate such that $M(r) = r$.
This is due to the fact that $M(r)$ is an increasing function of $r$. Therefore, $E$ and $t_b$ become functions of $M$.
For the initial time we choose the time of the last scattering surface: $t_1\simeq 3.77\times10^5 yr$. The initial density
profile should show a small over-density near the center imitating otherwise a FRW universe. Therefore, we add a Gaussian
peak to the FRW background density $\rho_b$. We know already that having an over-density in an otherwise homogeneous universe needs
a void to compensate for the extra mass within the over-density region. Therefore, to model this void we subtract a wider
gaussian peak:
\begin{equation}\label{initden}
\rho(R,t_1)=\rho_b(t_1)\left[\left(\delta_1e^{-\left(\frac{R}{R_0}\right)^2}- b_1\right) e^{-\left(\frac{R}{R_1}\right)^2}+1\right],
\end{equation}
where $\delta_1$ is the density contrast of the Gaussian peak, $R_0$ is the width of the Gaussian peak, and $R_1$ is the width of the
negative Gaussian profile. The mass compensation condition leads to an equation for $b_1$. For the final time we choose the time when
our null geodesy has the nearest distance to the center of our model structure. For instance if we set our lens at the redshift
$z\simeq0.2$ then $t_2\simeq6.98 Gyr$.\\
The density profile we choose for the final time is the universal halo density profile (NFW) \cite{NFW95} convolved with a negative
Gaussian profile to compensate the mass plus the background density at that time:
\begin{equation}
\rho(R,t_2) = \left(\rho_{NFW}-b_2 \rho_b(t_2)\right)e^{-\left(\frac{R}{R_2}\right)^2}+\rho_b(t_2),
\end{equation}
where
\begin{equation}
\rho_{NFW} = \rho_b(t_2)\frac{\delta_c}{\left(\frac{R}{R_s}\right)\left(1+\frac{R}{R_s}\right)^2}
\end{equation}
and
\begin{equation}
\delta_c = \frac{200}{3}\frac{c^3}{\ln(1+c)-\frac{c}{1+c}}.
\end{equation}
In our numerical calculation we will use typical NFW values $R_s=0.5Mpc$ and $c=5$ for a galaxy cluster. Note that at the
time $t_2$ a black hole singularity covered by an apparent horizon has already been evolved. Therefore, the NFW profile has
to be modified and a black hole mass greater than a minimum value has to be added to it at the center. This physical fact is reflected in a
shell crossing singularity if we take the familiar NFW profile similar to that assumed for the time $t_1$. The mass we have assumed
for this black hole singularity is about one thousandth of the mass up to the $R_s$ and equal to $5.66\times10^{11} M_{\odot}$.
Figs. \ref{emclnfw} and \ref{tnclnfw} shows the LTB functions $E$ and $t_b$ as a result of these boundary assumptions. Using these LTB
functions, the density profile of our model structure is obtained and depicted in Fig. \ref{nfwden}.
\begin{figure}[t]
\includegraphics[width = \columnwidth]{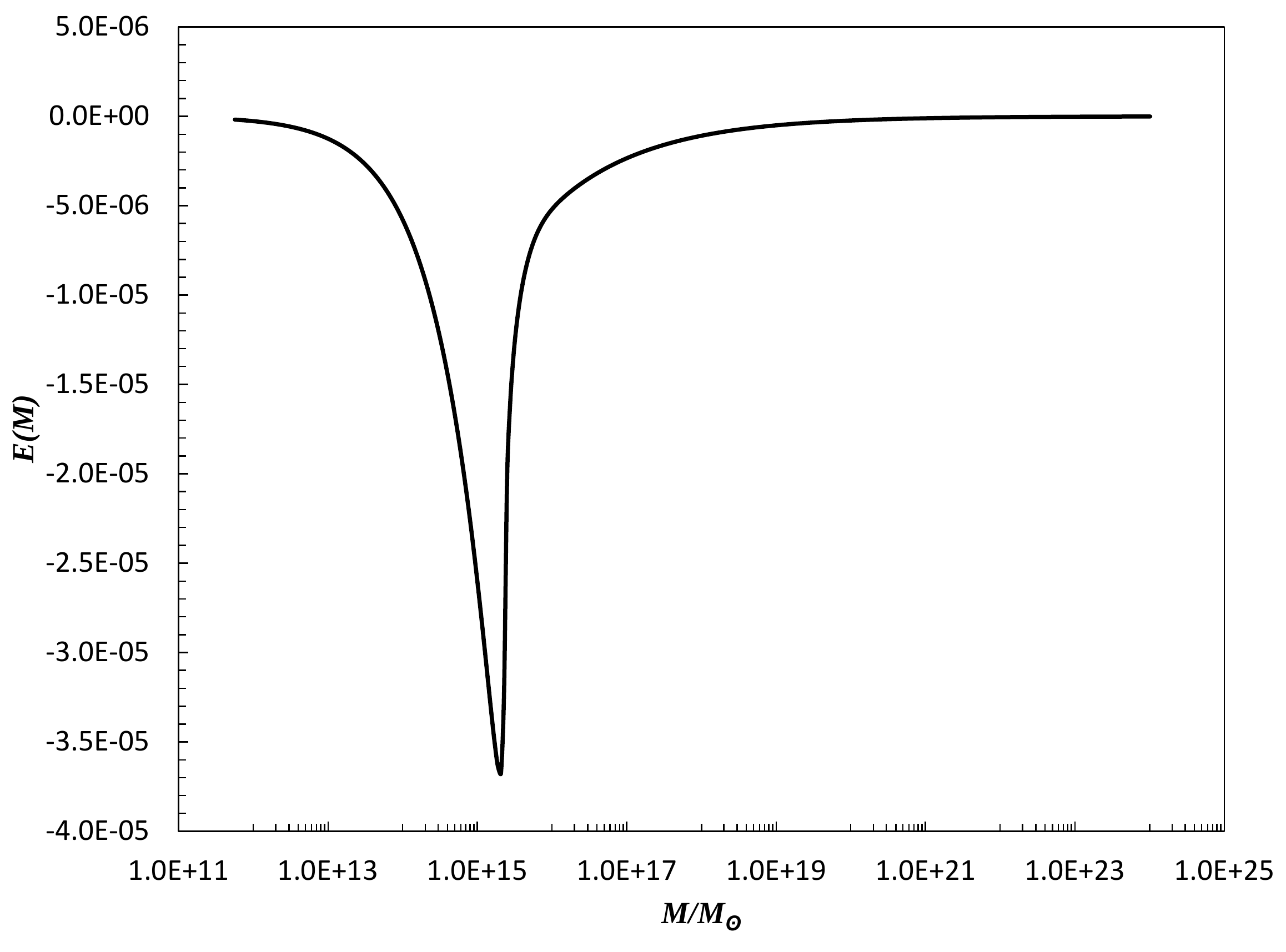}
\hspace*{10mm}\caption{\label{emclnfw}$E$ as a function of $M$ for a cluster with NFW density profile. $M$ is given in the unit of the Sun mass.}
\end{figure}
\begin{figure}[t]
\includegraphics[width = \columnwidth]{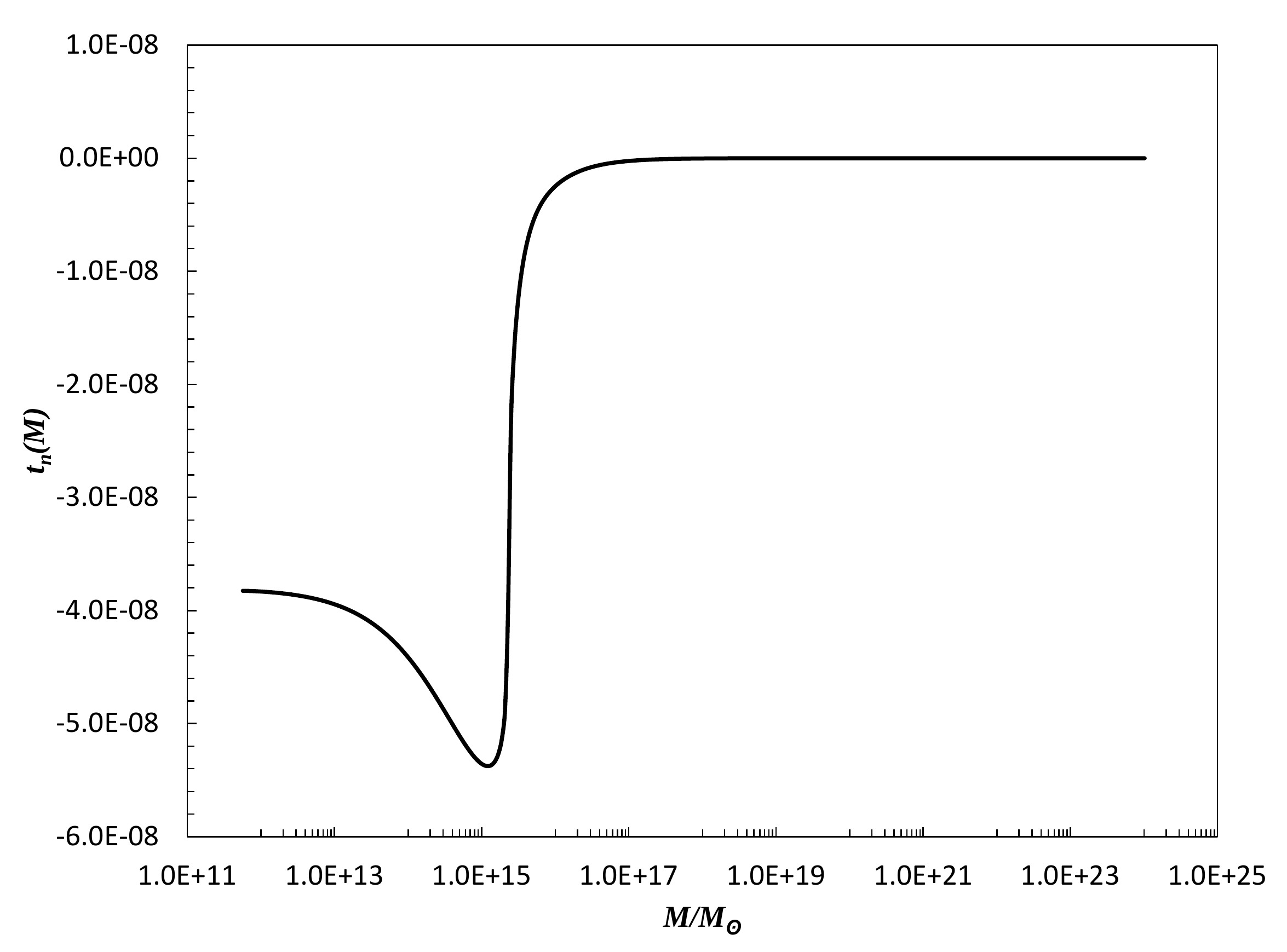}
\hspace*{10mm}\caption{\label{tnclnfw}$t_b$ as a function of $M$ for a cluster with NFW density profile. $t_b$ is given in the unit of $3.263 Gyr$.}
\end{figure}
\begin{figure}[t]
\includegraphics[width = \columnwidth]{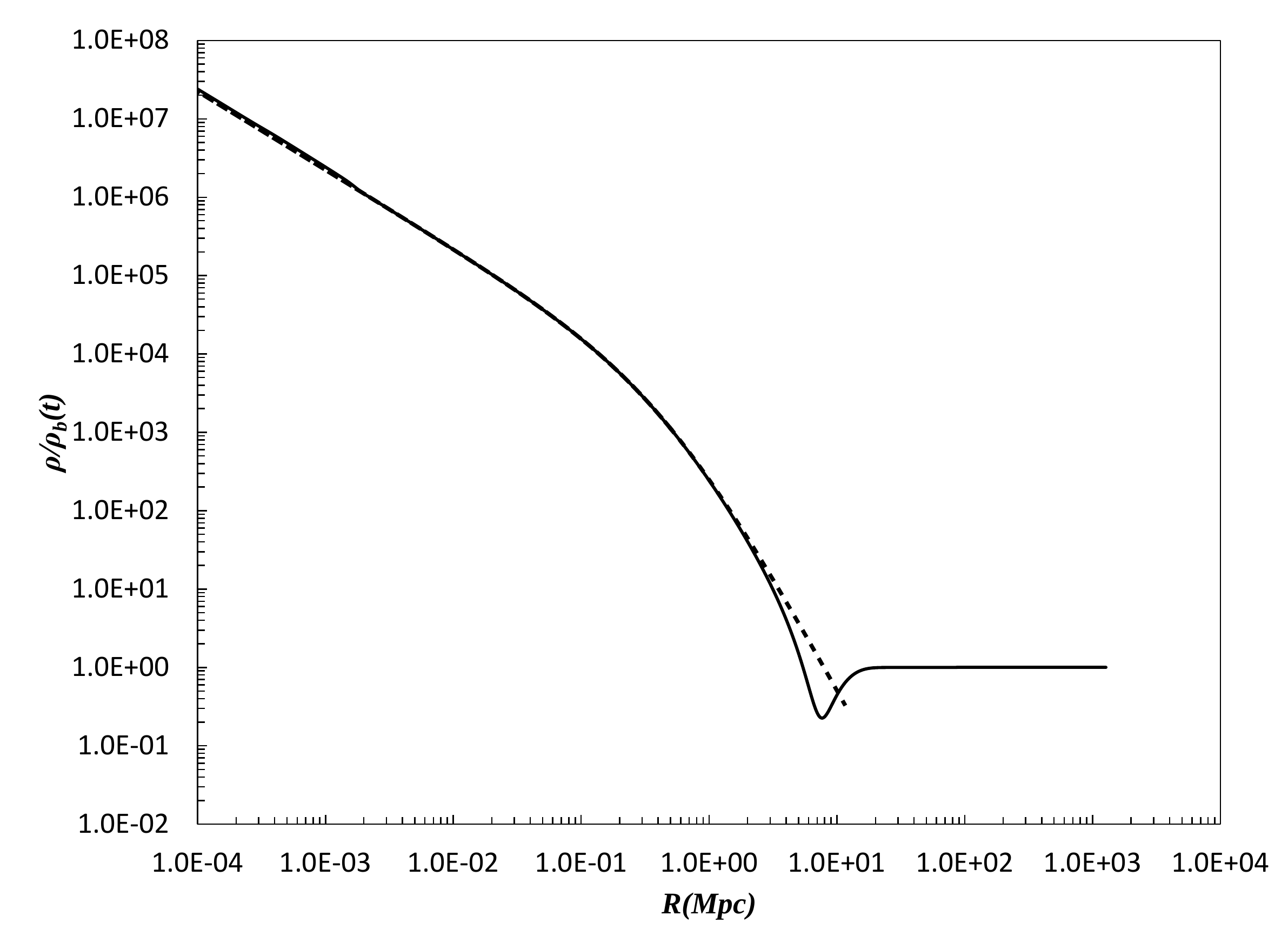}
\hspace*{10mm}\caption{\label{nfwden}Density profile for a cluster. The dot line corresponds to the familiar NFW profile and the solid line corresponds
to the modified NFW with a void.}
\end{figure}

To solve these equations we have to specify four initial conditions taking into account the light-like condition (\ref{null}). The freedom of choosing the affine parameter reduces the initial conditions to three. Now, the integration of the geodesics
happens by a backshooting procedure. Our initial conditions are taken to be the time of observation, distance of the observer
 to the lens expressed in terms of the redshift of the lens at the time of the observation, and angle between the line of sight
to the image of source and the line of sight to the lens ($\theta$ in Fig. \ref{lensdiag}):\\
\begin{equation}
\left.\tan\theta\right\vert_{O}=\left.\frac{R\frac{d\phi}{d\lambda}}{R'\frac{dr}{d\lambda}}\right\vert_{\text{Null}}.
\end{equation}
 The integration is done from the observer to the source at a specific redshift. Assuming there is no lens, the model reduces to a homogenous flat FRW universe and the geodesics are straight lines (in comoving coordinates) allowing us to determine the angle
between the source and the lens ($\beta$ in Fig. \ref{lensdiag}):
\begin{equation}
\tan\beta=\frac{\sin\phi_e}{\frac{r_o}{r_e}-\cos\phi_e},\\
\end{equation}
where $\phi_e$ is the $\widehat{OLS}$ angle, $r_o$ is the comoving distance of the observer, and $r_e$ is the comoving distance of the source from the center of coordinate system in the absence of lens at the time $t_e$. From the geodesic equations the $t_e$ is given by
\begin{equation}
\left(t_o^{\frac{1}{3}}-t_e^{\frac{1}{3}}\right)^2=\frac{1}{9}\left[\frac{R_o^2}{t_o^{\frac{4}{3}}}+ \frac{R_e^2}{t_e^{\frac{4}{3}}}-\frac{2R_o R_e}{t_o^{\frac{2}{3}}t_e^{\frac{2}{3}}}\cos\phi_e\right].
\end{equation}

\begin{figure}[t]
\includegraphics[width = \columnwidth]{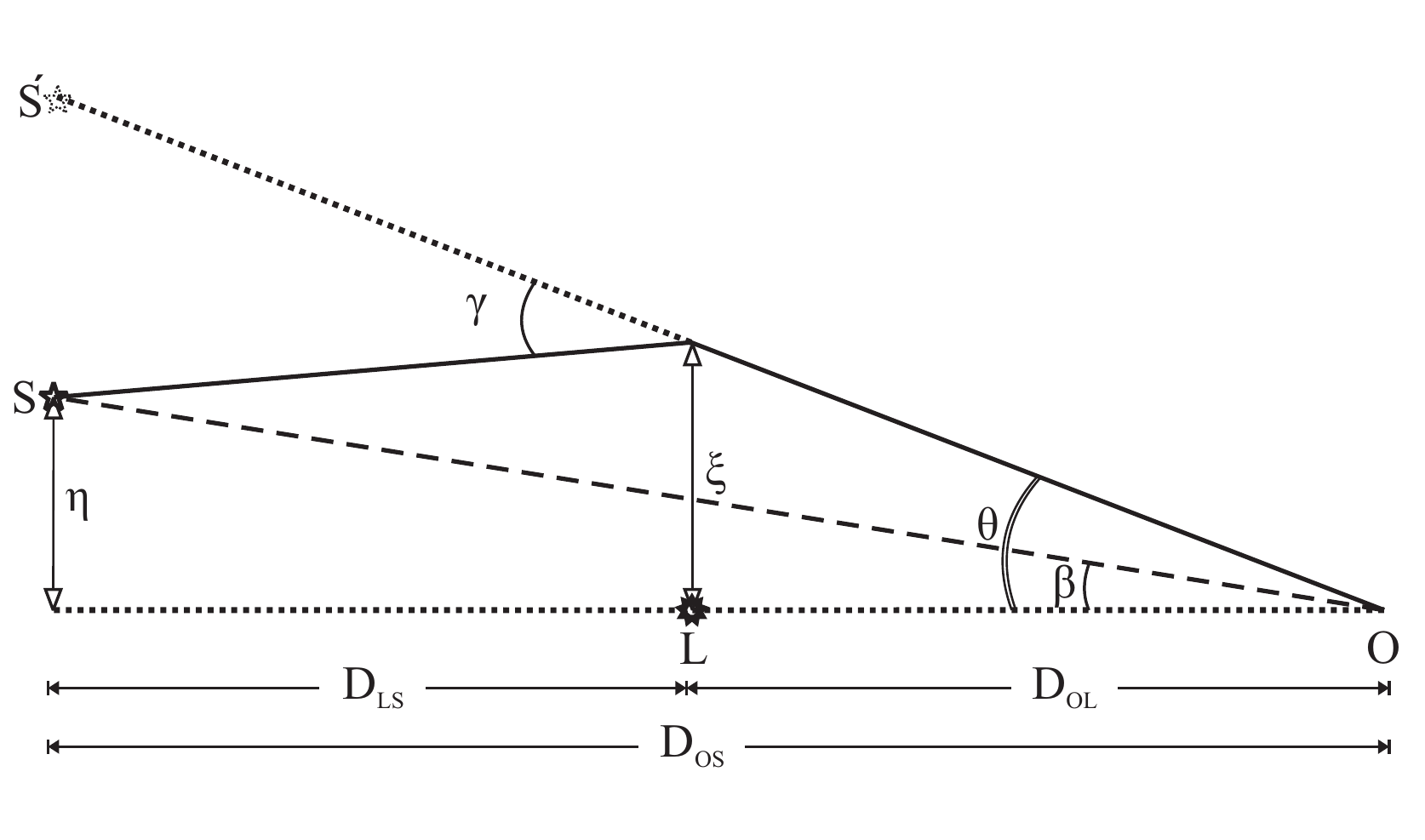}
\hspace*{10mm}\caption{\label{lensdiag}GL diagram: O is observer, S is source, S' is image in source plane, L is lens and $\gamma$ is deflection angle.}
\end{figure}

We then write the lens equation and determine the deflection angle $\gamma$:
\begin{equation}
\gamma=(\theta-\beta)\frac{D_{OS}}{D_{LS}},
\end{equation}
where we have assumed that the presence of the lens has not a significant effect on the distances and we may use the
corresponding FRW ones.\\

The validity of the numerical method chosen to integrate such complex system of partial differential equations is a
delicate issue. We first started with the familiar Runge-Kutta adaptive step size algorithm with proportional
and integral feedback (PI control) \cite{NR07} in which the step size is adjusted to keep local error under a suitable
threshold. We started with the so-called embedded Runge-Kutta of the rank 5(4). It turned out, however, that its accuracy
is too low. Therefore, we tried the rank 8(7) and then the rank 11(10) algorithm. The difference between these two last
ranks, however, turned out to be marginal and below one percent. Given the time-consuming rank 11(10) algorithm, we
preferred to use the rank 8(7) one. Now, as a fist test for the accuracy of this numerical method we tried the trivial
example of the LTB model, namely the FRW case, expecting a null result. The result was a non-negligible deflection angle
of the order of few milliarcseconds. Suspecting to face a numerical effect, and trying to understand the numerical
algorithm and the source of this numerical effect, we continued to calculate a more concrete and non-trivial LTB case.
The result for the rank 8(7) Runge-Kutta numerical method applied to a structure with a compact density profile did
agree with the thin lens approximation. However, in the case of a more diffuse density profile the result showed a
deflection angle up to an order of magnitude higher than the thin lens approximation. We did interpret this result
as a sign not to trust the Runge-Kutta method and turned to an alternative numerical method! \\

The root of this numerical deficiency could be due to the term $\frac{d\phi}{d\lambda}$ in our equations, which
is almost zero in the most part of the path of the light ray and changes suddenly to $\pi$ in the vicinity of the
lens. This is a well-known phenomenon in the numerical method of integrating differential equations called as "stiff" \cite{DV84}.
The characteristic property of such stiff equations is the presence of two quite different scales. In our case we have on one side
the cosmological distance scale of the source relative to the lens and the observer, and on the other side the scale of the structure or
the nearest distance of the ray to the lens. Realizing this stiffness property, we turned to the so-called semi-implicit
Rosenbrock method of the numerical integration of partial differential equations \cite{DV84,NR07}. As a first test we calculated
again the trivial case of a FRW model which gave an acceptable null result. We, therefore, decided to
integrate our geodesic equations using the semi-implicit Rosenbrock method instead of the Runge-Kutta one.\\

The null geodesics equations of our exact general relativistic structure model is now integrated using the
modified NFW density profile with a void before matching to the background FRW universe to obtain the deflection
angle. Note that the density in the NFW density profile is taken to be the oversdensity in an otherwise
FRW model, namely $\rho-\rho_b$. However, for the Th-L approximation we have used two different density profiles namely the familiar one and the modified one with a void before matching to the background density. In the case of  familiar NFW density profile without a void, the corresponding
equations can be integrated analytically to give the deviation angle \cite{bart96,keet02}:

\begin{eqnarray}
  \gamma(x) &=& {\frac{4M_{sing}}{x R_s}}+{16\pi\rho_b \delta_c \frac{R_s^2}{x}}
  {\left( \log{\frac{x}{2}} + F(x)\right)}\label{nfwgamma}\\
  F(x)&=&\left\{\begin{array}{lr}
               \frac{\textrm{arctanh}({\sqrt{1-x^2}})}{\sqrt{1-x^2}}& x<1 \\
               1 & x=1 \\
               \frac{\arctan({\sqrt{x^2-1}})}{\sqrt{x^2-1}}& x>1
             \end{array}\right.
\end{eqnarray}
Assuming the same modified NFW profile as in general relativistic case for the Th-L approximation
we have also calculated the deflection angle applying the lens equation \cite{GLenses}
\begin{equation}\label{lenseqn}
    \theta-\beta=\frac{D_{LS}}{D_{OL}D_{OS}}\frac{d\Psi(\theta)}{d\theta},
\end{equation}
where $\Psi$ is the lens potential.

\begin{figure}[t]
\includegraphics[width = \columnwidth]{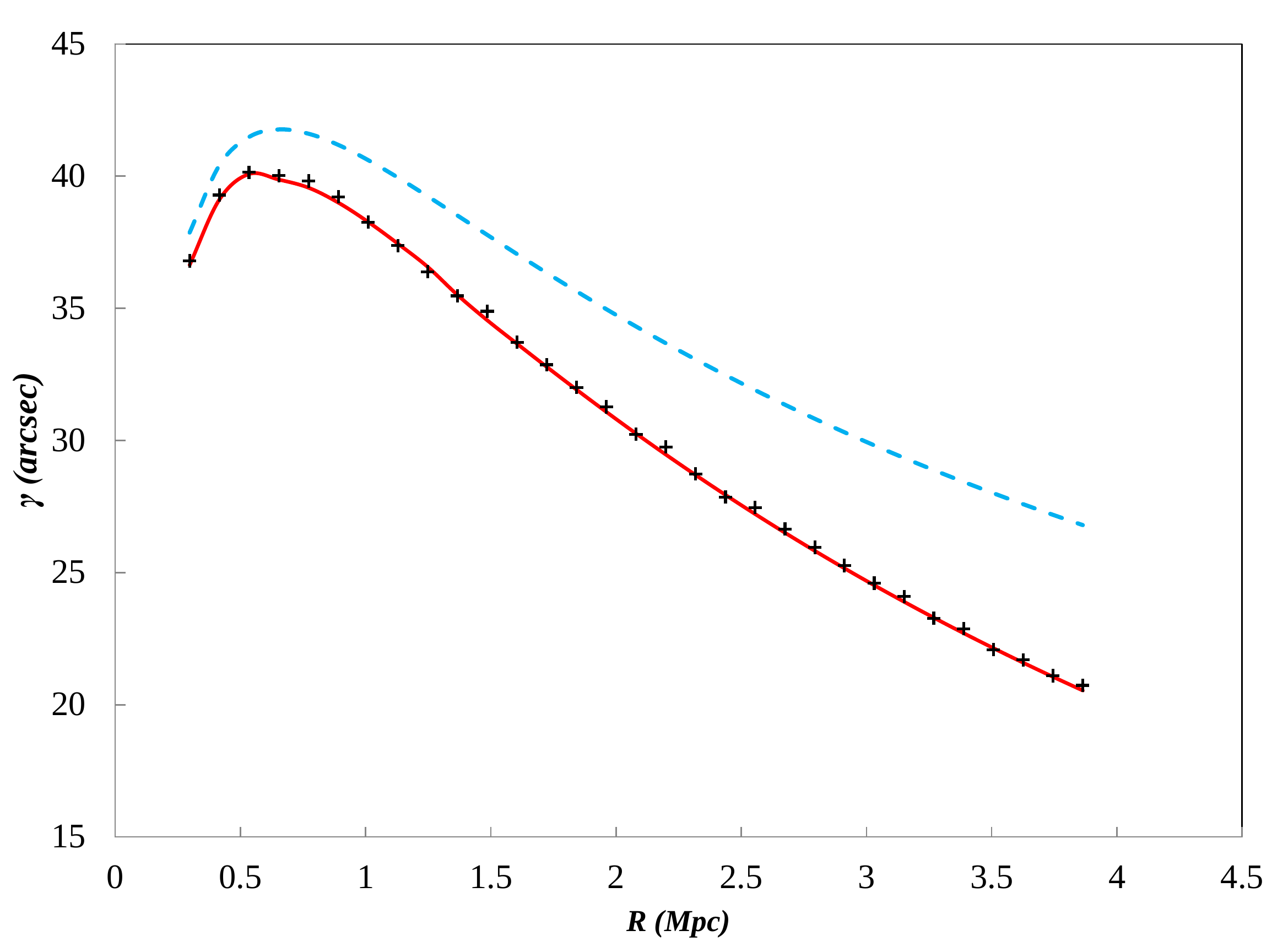}
\hspace*{10mm}\caption{\label{nfwdev}Deviation angle for three cases: the general relativistic result is indicated by plus points; the thin lens approximation using our modified NFW is shown by the continuous line; and the dashed line is for the familiar NFW profile without the void (formula (\ref{nfwgamma})).}
\end{figure}

The result for the three cases, the exact general relativistic model with our modified NFW profile, thin lens approximation using the
modified NFW with void, and the thin lens approximation using the familiar NFW without a void is depicted in Fig. \ref{nfwdev}. Obviously
the two cases of the thin lens approximation with the modified NFW density profile including the void and the LTB exact method almost coincide.
\begin{figure}[t]
\includegraphics[width = \columnwidth]{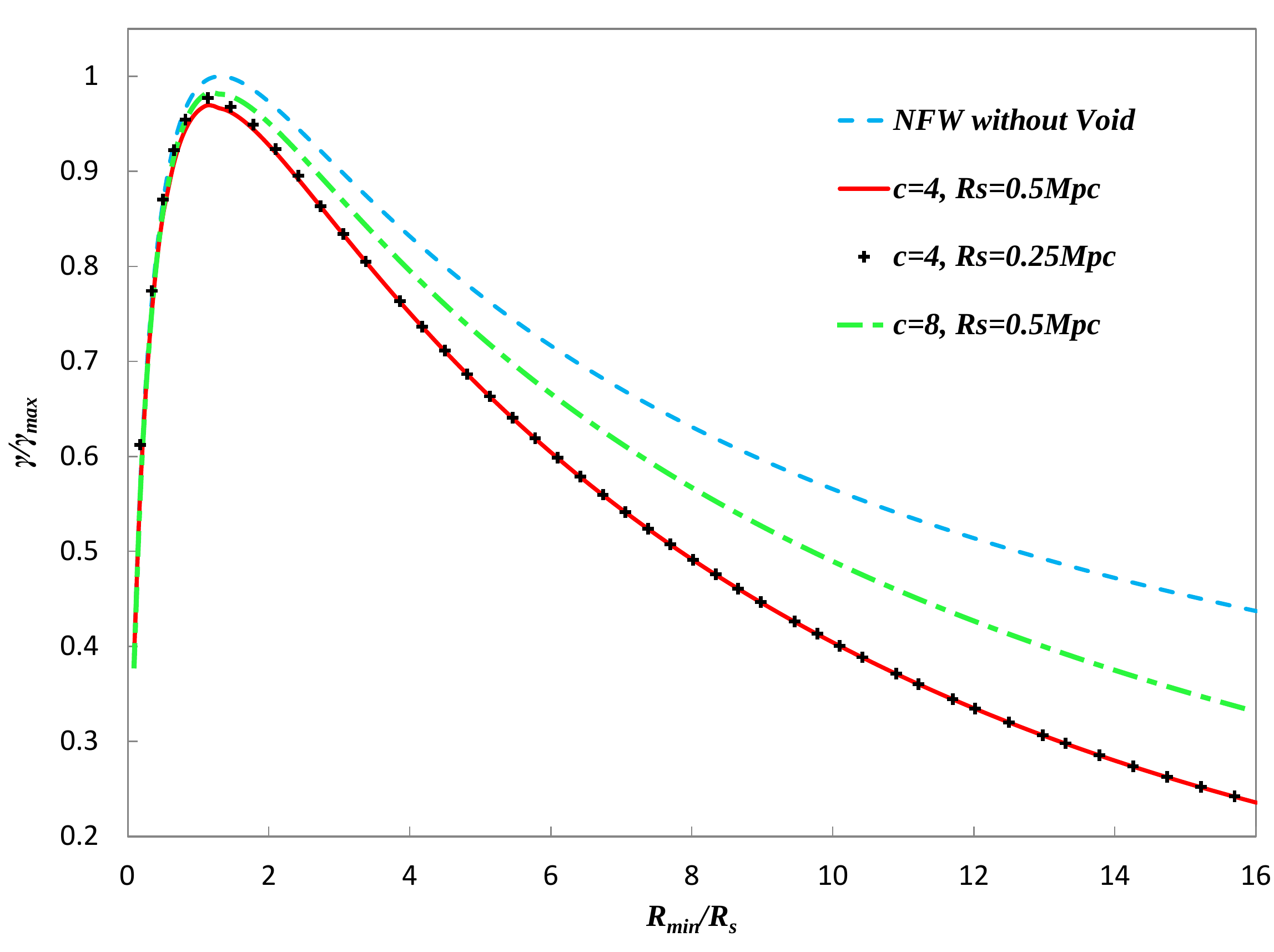}
\hspace*{10mm}\caption{\label{diffcrs}Deviation angle for NFW density profiles with different parameters. Horizontal axis is normalized to $R_s$ and vertical axis is normalized to the maximum of the deflection angle in each case. Dash line is for NFW model without void (formula (\ref{nfwgamma})). }
\end{figure}
The thin lens approximation with the familiar density profile without a void, however, differ from the exact LTB model. The difference
in the deviation angle can be more than 30 percent depending on the impact parameter. The difference between the exact general
relativistic LTB model and the thin lens approximation is due to the absence of the void in the familiar NFW profile used in the
literature. To see the implications of the NFW parameters in this difference we have also calculated the deviation
angle for different NFW profiles, with and without void. The result is depicted in the Fig. \ref{nfwgamma}. We see again
that the Th-L approximation using different modified NFW profiles including a void almost coincide with the exact LTB model.
Models with the NFW profiles without void, however, differ substantially from the exact model. The difference is higher
the bigger the $c_s$ parameter is, i.e. the less the concentration of the density of structure is. \\
We, therefore, conclude that by interpreting astrophysical data of gravitational lensing by clusters using a familiar
NFW density profile without a void we are deviating from the exact result and the Th-L approximation is no longer valid.
The Th-L approximation may, however, be considered as precise enough if one modify the density profile and add the
corresponding void to it, as require by general relativity for a quasi-local structure. The detail of the void, such as
its density contrast,its depth and length, depends on the detail of the model and the deviation from the familiar NFW may
even be much higher for other choices. Also note that the effect of the void is higher for larger impact parameter. In
the case of strongly lensed objects in astrophysical applications we are usually faced with small impact parameter
where this effect is negligible. For example in the case of Abell 2261 cluster ($z=0.225$) with many strong lensing
arcs, D. Coe et al. \cite{coe} have assigned $c_s=6.2\pm 0.3$ and $M_{vir}=2.2\pm 0.2\times10^{15}M_\odot$. The exact
general relativistic results according to our model would lead to $c_s=6.23$ and $M_{vir}=2.23\times10^{15}M_\odot$.
In the case of weak lensing, however, we expect this effect to have significant impact on the cosmological parameters.
Work in this direction is in progress.

\end{document}